\documentclass[aps,twocolumn,groupedaddress,showpacs]{revtex4}
\usepackage{graphicx,amssymb,amsbsy,color}
\begin{document}

\title{Interplay between antiferromagnetic order and spin polarization
in ferromagnetic metal/electron-doped cuprate superconductor junctions}
\author{Pok-Man Chiu,$^{1}$ C. S. Liu,$^{2,1}$ and W. C. Wu$^{1}$}
\affiliation{$^1$Department of Physics, National Taiwan Normal
Univesity, Taipei 11650, Taiwan\\
$^2$Department of Physics, Yanshan University, Qinhuangdao 066004,
China}
\date{\today}

\begin{abstract}
Recently we proposed a theory of point-contact spectroscopy and
argued that the splitting of zero-bias conductance peak (ZBCP) in
electron-doped cuprate superconductor point-contact spectroscopy
is due to the coexistence of antiferromagnetic (AF) and $d$-wave
superconducting orders [Phys.~Rev.~B {\bf 76}, 220504(R) (2007)].
Here we extend the theory to study the tunneling in the
ferromagnetic metal/electron-doped cuprate superconductor
(FM/EDSC) junctions. In addition to the AF order, the effects of
spin polarization, Fermi-wave vector mismatch (FWM) between the FM
and EDSC regions, and effective barrier are investigated. It is
shown that there exits midgap surface state (MSS) contribution to
the conductance to which Andreev reflections are largely modified
due to the interplay between the exchange field of ferromagnetic
metal and the AF order in EDSC. Low-energy anomalous conductance
enhancement can occur which could further test the existence of AF
order in EDSC. Finally, we propose a more accurate formula in
determining the spin polarization value in combination with the
point-contact conductance data.
\end{abstract}

 \maketitle

\section{Introduction}
Using point contact technique to measure the spin polarization in
ferromagnetic metal/conventional superconductor (FM/CS) junctions
was pioneeringly done by Soulen {\em et al.} \cite{Soulen98} and
Upadhyay {\em et al.} \cite{Upadhyay98} in 1998. Their works
showed that determining the spin polarization at Fermi surface is
essentially not an easy task. That leads to some definitions of
spin polarization such as ``tunneling polarization" proposed by
Tedrow and Meservey \cite{Tedrow94} and ``point-contact
polarization" proposed by Soulen {\em et al.} \cite{Soulen98}. One
year later, Zhu {\em et al.} \cite{Zhu99,Zhu00} and Kashiwaya {\em
et al.} \cite{Kashiwaya99} have utilized the ideas to study the
spin-polarized quasiparticle transport in ferromagnet/$d$-wave
superconductor junctions. Zhu {\em et al.} \cite{Zhu99,Zhu00}
predicted that conductance resonances occur in a
normal-metal-ferromagnet/$d$-wave superconductor junction and in a
following paper, they further studied the junctions by solving the
Bogoliubov-de Gennes (BdG) equations within an extended Hubbard
model which included the proximity effect, the spin-flip
interfacial scattering at the interface, and the local magnetic
moment. They have reported that the proximity can induce order
parameter oscillation in the ferromagnetic region. In contrast,
Kashiwaya {\em et al.} \cite{Kashiwaya99} focused on the spin
current and spin filtering effects at the magnetic interface. In
the works of Zutic and Valls \cite{Zutic99,Zutic00}, they first
considered the effect of Fermi-wave vector mismatch (FWM) and have
pointed out that if one neglects FWM, the effect of spin
polarization invariably leads to the suppression of Andreev
reflection (AR). Among many other junction studies, Dong {\em et
al.} \cite{Dong01} studied a little different junction which forms
a four layer sandwich, {\em i.e.}, FM/I/$d+ is$/$d$-wave
junctions, by taking into account the roughness of the interfacial
barrier and broken time-reversal symmetry states.

The pioneering works of Soulen {\em et al.} and Upadhyay {\em et
al.} have inspired several experimental studies
\cite{Ji01,Strijkers01,Kant02,Raychaudhuri03,Perez-Willard04,
Woods04,Mukhopadhyay07,Chalsani07} as well.  Especially normal and
ferromagnetic metal/conventional superconductor or $s$-wave
superconductor (FM/$s$-wave SC) junctions have been intensely
studied experimentally and theoretical modelings
(Blonder-Tinkham-Klapwijk (BTK) formula \cite{Blonder82} or its
extension) had a good fitting with the conductance data. Recently
Linder and Sudb{\o} \cite{Linder07} presented a theoretical study
of FM/$s$-wave SC junction that investigated the possibility of
induced triplet pairing state in the ferromagnetic metal side.
They have also used the BTK approach but allowed for arbitrary
magnetization strength and direction in the ferromagnet, arbitrary
spin-active barrier, arbitrary FWM, and different effective masses
in the two sides of the junction. As is expected, there is no
retroreflection process when an exchange field is present.
However, they pointed out that retroreflection can occur under
some conditions \cite{Linder07}.

If one replaces the conventional superconductor by the
high-temperature or $d$-wave superconductor into the junction, it
will occur several novel phenomena due to its $d$-wave pairing
symmetry, complex band structure, and rich magnetic properties. Of
particular interest, in the electron-doped side of cuprate
superconductors (EDSC), it is strongly suggested that
antiferromagnetic (AF) order may coexist with the $d$-wave
superconducting order, especially in the underdoped and
optimally-doped regimes \cite{Liu07}. In this paper, we shall
explore the possible novel phenomena in the FM/EDSC junction case,
taking into account the interplay between antiferromagnetic order
and spin polarization. The ideas and models developed in FM/CS
junctions in the literature will be applied to the current FM/EDSC
junction cases.

This paper is organized as follows. In Sec.~II, the basic
formulation is given. We set up the condition of the junction and
generalize the BdG equations to include AF order parameter. As the
formal process, we utilize WKBJ approximation to obtain the more
simple Andreev-like equations, which are then solved to determine
the four spin-dependent reflection coefficients (detailed
derivations are given in Appendix~\ref{appendix:A}). Formulas of
charge and spin conductances are derived . Sec.~III are our main
results and discussions. In Sec.~III.A, the condition of midgap
surface states was derived (details are given in
Appendix~\ref{appendix:B}). In Sec.~III.B, the effect of FWM was
studied. In Secs.~III.C and III.D, we discuss the effects of
spin-polarization and generalized effective barrier, respectively.
It is shown that anomalous conductance enhancement can occur at
low energies which could provide a further test for the existence
of AF order in EDSC. In Sec.~III.E, a more general formula for
determining the spin polarization is proposed in terms of the
experimental zero-bias conductance data. Finally in Sec.~IV, a
brief conclusion is given.

\section{Formalism}
Our formulation is given based on the following assumptions. We
consider a point contact or planar FM/I/EDSC junction where the
superconductor overlayer is coated with a clean, size-quantized,
ferromagnetic-metal overlayer of thickness $d$, that is much
shorter than the mean free path $l$ of normal electrons. The
interface is assumed to be perfectly flat and infinitely large.
Considering $l\rightarrow \infty$ limit, the discontinuity of all
parameters at the interface can be neglected, except for the SC
order parameter to which the proximity effect is ignored
\cite{Hu94}. When SC and AF orders coexist, quasiparticle (QP)
excitations of an inhomogeneous superconductor can have a coupled
electron-hole character associated with the coupled $\mathbf{k}$
and ${\bf k}+{\bf Q}$ [$\mathbf{Q}=(\pi ,\pi )$] subspaces. QP
states are thus governed by the generalized BdG equations
\cite{Gennes66,Liu07}
\begin{eqnarray}
 E u_{1\sigma} ({\bf{x}}) &=& \hat H_\sigma u_{1\sigma} ({\bf{x}}) +
 \int {d{\bf{x}^\prime}\Delta({\bf{s}},{\bf{r}})v_{1\bar\sigma} ({\bf{x}^\prime}) +
 \Phi u_{2\sigma} ({\bf{x}})} \nonumber \\
 E v_{1\bar\sigma} ({\bf{x}}) &=& \int {d{\bf{x}^\prime}\Delta^*
({\bf{s}},{\bf{r}})u_{1\sigma} ({\bf{x}^\prime}) - \hat H_\sigma
v_{1\bar\sigma}
 ({\bf{x}}) + \Phi v_{2\bar\sigma} ({\bf{x}})} \nonumber \\
 E u_{2\sigma} ({\bf{x}}) &=& \Phi u_{1\sigma} ({\bf{x}})+
 \hat H_\sigma u_{2\sigma} ({\bf{x}}) -
 \int {d{\bf{x}^\prime}\Delta ({\bf{s}},{\bf{r}})v_{2\bar\sigma} ({\bf{x}^\prime})}
 \nonumber \\
 E v_{2\bar\sigma} ({\bf{x}}) &=& \Phi v_{1\bar\sigma}
  ({\bf{x}})-\int {d{\bf{x}^\prime}\Delta^*({\bf{s}},{\bf{r}})u_{2\sigma} ({\bf{x}^\prime})
  - \hat H_\sigma v_{2\bar\sigma} ({\bf{x}})},\nonumber\\
\label{eq:BdG}
\end{eqnarray}
where $ {\bf{s}} \equiv {\bf{x}}  - {\bf{x}^\prime}$, $ {\bf{r}}
\equiv ({\bf{x}} + {\bf{x}^\prime} )/2 $, $\hat H_{_\sigma}\equiv
-\hbar ^{2}\nabla ^{2}/2m-E_F^{F,S}-\sigma J$ with $ J $ the
exchange energy and $\sigma = \uparrow$ ($\downarrow$) for up
(down) spin ($\bar{\sigma}=-\sigma$), and $\Phi $ is the AF order
parameter. $\Delta ({\bf{s}},{\bf{r}})$ is the Cooper pair order
parameter in terms of relative and center-of-mass coordinates. In
the FM region, we define $E_F^F \equiv \hbar ^2 q_{F}^2 /2m =
(\hbar ^2 q_{F \uparrow }^2 /2m + \hbar ^2 q_{F \downarrow }^2
/2m)/2$ as the spin averaged value. It differs from the value in
the superconductor, $E_F^S \equiv \hbar ^2 k_{F}^2 /2m$, to which
a FWM can occur between the FM and EDSC regions \cite{Zutic00}. In
(\ref{eq:BdG}), the wave functions $u_{1}$ and $v_{1}$ are
considered related to the $\mathbf{k}$ subspace, while $u_{2}$ and
$v_{2}$ are related to the ${\bf k}+{\bf Q}$ subspace. Comparing
with the first and second lines of Eq.~(\ref{eq:BdG}), minus signs
associated with the $\Delta({\bf{s}},{\bf{r}})$ term in the third
and fourth lines occur due to the symmetry requirement, $\Delta
({\bf k}+{\bf Q}) = - \Delta ({\bf{k}})$, for a $d_{x^2-y^2}$-wave
superconductor in ${\bf k}$ space. At Fermi level, the
$d_{x^2-y^2}$-wave SC gap is $\Delta ({\bf{\hat k}}_F)\equiv
\Delta _{0}\sin 2\theta$ with $\Delta_0$ the gap magnitude and
$\theta$ the azimuthal angle relative to the $x$-axis.

\begin{figure}[ptb]
\begin{center}
\includegraphics[
width=8.5cm ]{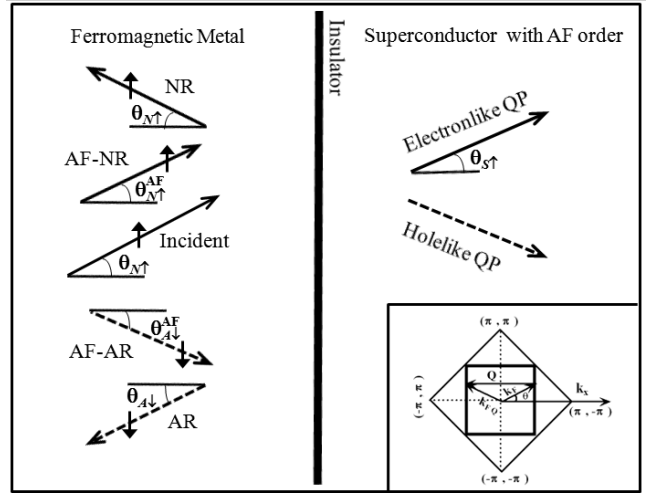}
\end{center}
\par \vspace{-0.5cm} \caption{Schematic plot showing all possible
reflection and transmission processes for an up-spin electron
incident into the FM/I/EDSC junction. An AF order is assumed to
exist in the EDSC. For convenience for a $d$-wave superconductor,
$\mathbf{k_{x}}$ axis is chosen to be along the [110] direction.
The right-bottom inset shows a given Fermi wave vector $%
\mathbf{k}_{F}=\left( k_{F} ,k_{y},k_{z}\right) $ and its coupled
AF wave vector
$\mathbf{k}_{F}+\mathbf{Q}\equiv\mathbf{k}_{F\mathbf{Q}}=\left(
-k_{F},k_{y},k_{z}\right)$. Both vectors are tied to the Fermi
surface, which is approximated by a square (thick line). NR, AR,
AF-NR, and AF-AR stand for normal reflection, Andreev reflection,
antiferromagnetic-normal reflection, and antiferromagnetic-Andreev
reflection respectively. Their corresponding reflection angles are
also shown. For the case of an incident down-spin electron, all
spin indices just reverse.} \label{fig1}
\end{figure}

In a $d$-wave superconductor, it's useful to consider a junction
to which the superconductor surface is allied along the [110]
direction. A thin insulating layer exists between the
ferromagnetic metal and the superconductor (see Fig.~\ref{fig1})
to which the barrier potential is assumed to take a delta
function, $V(x)=H\delta \left( x\right)$. Considering that an
up-spin electron is injected into the FM/I/EDSC junction from the
ferromagnetic metal side, there are four possible reflections as
follows: (a) Normal reflection (NR): reflected as electrons. (b)
Andreev reflection (AR): reflected as holes, due to electron and
hole coupling in the $\mathbf{k}$ subspace. (c)
Antiferromagnetic-Normal reflection (AF-NR): reflected as
electrons, due to the coupling of $\mathbf{k}$ and ${\bf k}+{\bf
Q}$ subspaces. (d) Antiferromagnetic-Andreev reflection (AF-AR):
reflected as holes, due to electron and hole coupling in the ${\bf
k}+{\bf Q}$ subspace (see Fig.~\ref{fig1}).

In addition to the effect of AF order, AR is largely modified due
to the exchange field of ferromagnetic metal when electron is not
normally incident into the EDSC region. Owing to the momentum
conserved parallel to the interface, Snell's law
\cite{Jackson75,Zutic00,Kashiwaya99} requires that
\begin{eqnarray}
 q_{F\sigma } \sin \theta_{N\sigma}=
 q_{F\bar \sigma } \sin \theta_{A\bar \sigma }
 =& k_{F}\sin \theta _{S\sigma },
\end{eqnarray}
where $\theta_{N\sigma}$, $\theta_{A\bar \sigma }$, and $\theta
_{S\sigma }$ are the angles of NR, AR, and transmission into the
SC respectively (see Fig.~\ref{fig1}). Incident angle
$\theta_{N\sigma}$ is typically {\em not} equal to the AR angle
$\theta_{A\bar \sigma }$ except when $J=0$ or for normal
incidence. Assuming that there is no FWM and $q_{F \downarrow
}<k_F<q_{F \uparrow }$, ranges of six normal and Andreev
reflection angles are $0 < \theta _{N \uparrow } < \sin ^{ - 1}
(k_{F} /q_{F \uparrow } )\equiv\theta _{c2} $, $ 0 < \theta _{A
\uparrow } < \sin ^{ - 1} (q_{F \downarrow } /q_{F \uparrow }
)\equiv\theta _{c1}$, and $0 < \theta _{S \uparrow},\theta _{S
\downarrow} < \sin ^{ - 1} (q_{F \downarrow } /k_{F} )$, while
$\theta _{A \downarrow }$ and $\theta _{N \downarrow}$ can be any
angles. For AF reflections, the angles $\theta _{A\sigma}^{\rm
AF}=\pi-\theta _{A\sigma }$ and $\theta_{N\sigma }^{\rm AF}=\theta
_{N\sigma }$ respectively. It is noted that when $\theta _{N
\uparrow }$ is within the range $\theta _{c1} < \theta _{N
\uparrow } < \theta _{c2}$, $x$ component of the wave vector,
$\sqrt {q_{F \downarrow }^2 - k_{F}^2 \sin ^2 \theta _{S \uparrow
} }$, becomes purely imaginary for the AR process
\cite{Zutic00,Kashiwaya99}. Although spin down electron as a
propagating wave is impossible for AR, it can still transmit into
the superconductor side.

As emphasized by Kashiwaya {\em et al.} \cite{Kashiwaya99}, one
can define two types of conductance in a FM, namely the {\em
charge} and {\em spin} conductances.  As a matter of fact, the
normalized angle and spin dependent tunneling charge conductance
is given by
\begin{eqnarray}
C_{q \sigma }  = 1 - \left| {R_{N \sigma} } \right|^2  +
a_{\sigma}\left| {R_{A \bar \sigma}} \right|^2  + \left|
{R^{AF}_{N \sigma } } \right|^2  - a_{\sigma}\left| {R^{AF}_{A\bar
\sigma }} \right|^2, \label{eq:Cq}
\end{eqnarray}
where $a_{\downarrow}\equiv 1$ and $a_{\uparrow}\equiv L_
\downarrow \lambda _{2 \downarrow } /L_ \uparrow  \lambda _{1
\uparrow}$ with $\lambda _{1\uparrow} = \cos \theta _{N\uparrow}
/\cos \theta _{S\uparrow} $, $ \lambda _{2\downarrow} = \cos
\theta _{A\downarrow} /\cos \theta _{S\uparrow}$, and $L_\sigma
=\sqrt {(q_F /k_F)(1- \sigma J/E_F^F)}$. Detailed derivations of
all four reflection coefficients ($R_{N \sigma}$, $R_{A \bar
\sigma}$, $R^{AF}_{N \sigma }$, and $R^{AF}_{A\bar \sigma }$) are
given in Appendix~\ref{appendix:A}. Similarly, the normalized
angle and spin dependent spin conductance is given by
\begin{eqnarray}
C_{s \sigma }  = 1 - \left| {R_{N \sigma } } \right|^2  -
a_{\sigma}\left| {R_{A \bar \sigma } } \right|^2  + \left|
{R^{AF}_{N\sigma }} \right|^2  - a_{\sigma}\left| {R^{AF}_{A\bar
\sigma}} \right|^2. \label{eq:Cs}\end{eqnarray} Comparing with the
results of charge conductance in (\ref{eq:Cq}), due to the spin
imbalance induced by the exchange field, different sign of
$R_{A\bar{\sigma}}$ terms occurs in the spin conductances.
Consequently normalized total charge (spin) conductance is given
by
\begin{eqnarray}
G_{q(s)}(E) = G_{q(s)\uparrow}(E)\pm G_{q(s)\downarrow}(E),
\end{eqnarray}
where $+$ ($-$) sign is for charge (spin) channel and
\begin{eqnarray}
G_{q(s)\sigma }(E)  = \frac{1}{{G^N_{q(s)} }}\int_\alpha ^\beta
{d\theta _{N\sigma} \cos \theta _{N\sigma} C_{q(s)\sigma}
(E,\theta _{N\sigma} )P_\sigma}. \label{eq:P1}
\end{eqnarray}
The lower and upper integration limits of $\alpha$ and $\beta$ are
restricted by Snell's law (as discussed before) or experimental
setup. In practice, integration over two separate ranges of
incident angle, {\em i.e.}, $0 < \left| {\theta _{N\sigma} }
\right| < \theta _{c1}$ and $\theta _{c1}<\left| {\theta
_{N\sigma} } \right|<\theta _{c2}$ should be carried and results
are added up to the total conductance. In (\ref{eq:P1}), the
normal-state charge (spin) conductance
\begin{eqnarray} G^N_{q(s)}  = \int_{ - \pi /2}^{\pi /2} {d\theta
_{N\sigma} \cos \theta _{N\sigma} [C_{N \uparrow } P_ \uparrow \pm
C_{N \downarrow } P_ \downarrow  ]}, \label{eq:P2}
\end{eqnarray}
where
\begin{eqnarray}
C_{N\sigma } (\theta _{N\sigma} ) = \frac{{4\lambda _1 L_\sigma
}}{{\left| {1 + \lambda _1 L_\sigma   + 2iZ} \right|^2 }}
\end{eqnarray}
with $Z=mH/\hbar ^2 k_{F}$  the barrier. In both (\ref{eq:P1}) and
(\ref{eq:P2}), we have introduced a factor $ P_\sigma =(
E_F^F+\sigma J )/2E_F^F$ which can be interpreted as the
probability of spin-$\sigma$ incident electron as a function of
the exchange energy \cite{Kashiwaya99,Zutic00,Linder07}. When
$J=0$, $P_\uparrow=P_\downarrow =1/2$.

In addition to the conductances, the normalized {\em total} charge
(spin) current can be given by
\begin{eqnarray}
\begin{array}{l}
I_{q(s)} = I_{q(s)\uparrow} \pm I_ {q(s)\downarrow},  \\
 \end{array}
\end{eqnarray}
where
\begin{eqnarray}
I_{q(s)\sigma } &=& \frac{1}{I^N_{q(s)}} \int_{ - \infty }^\infty
{dE} \int_\alpha ^\beta  d \theta _{N\sigma} \cos \theta
_{N\sigma} C_{q(s)\sigma } (E,\theta _{N\sigma} )P_\sigma
q_{F\sigma }\nonumber\\
\end{eqnarray}
with
\begin{eqnarray} {I^N_{q(s)}}  &=& \int_{ - \infty }^\infty  {dE} \int_{ - \pi
/2}^{\pi /2} {d\theta _{N\sigma} \cos \theta _{N\sigma} [C_{N
\uparrow } P_ \uparrow  q_{F \uparrow }\pm C_{N \downarrow } P_
\downarrow q_{F \downarrow } ]}. \nonumber\\
\end{eqnarray}
Charge and spin currents and their conversion are important probes
for spin-related phenomena such as those in spin Hall effect.
%

\section{Results and Discussions}

Both charge and spin conductances are important probes for
tunneling in spin-polarized junctions. In this paper, we will
focus on the charge conductance however. Moreover, for simplicity,
all the results presented are for normal incidence
($\theta_{N\sigma}=0$).

\subsection{Midgap Surface States}

Detailed derivations of the midgap surface states (MSS) in the
current FM/EDSC junction are give in Appendix~\ref{appendix:B}.
Basically it is an extension of Hu's \cite{Hu94} and Liu and Wu's
\cite{Liu07} works. The boundary condition that leads to the MSS
is the wave function $\psi_{N\sigma}(x=-d)=0$ for a free boundary
at $x=-d$. Consequently one obtains the following condition for
the MSS (see Appendix~\ref{appendix:B}):
\begin{eqnarray}
e^{ - 2ik_{1\sigma} d} E_{  + }  + e^{2ik_{1\sigma} d} E_{  - }  =
2\Phi, \label{surface bound states}
\end{eqnarray}
where $E_{ \pm } \equiv E \pm \varepsilon^\prime_\sigma$ with $
\varepsilon^\prime_\sigma   = \sqrt {(E + \sigma J)^2  - \Delta ^2
- \Phi ^2 }$ incident spin-$\sigma$ electron is assumed to have
wave vector $k_{1\sigma}$ along the $x$ direction.

In case of $J=0$, the result is reduced to our previous case
without spin polarization \cite{Liu07}. In case of $J=\Phi=0$, the
result is reduced to Hu's case \cite{Hu94}, {\em i.e.},
\begin{eqnarray}
e^{4ik_1 d}  =  -
\frac{{E+{\varepsilon^\prime}}}{{E-{\varepsilon^\prime}}},
\end{eqnarray}
where $\varepsilon^\prime\equiv \sqrt {E^2  - \Delta ^2}$. The
most crucial result of the above is that there exists a
zero-energy state which is
responsible for the ZBCP widely observed in hole-doped $d_{x^{2}-y^{2}}$%
-wave cuprate superconductors \cite{Hu94}. When $J=0$ but $\Phi
\neq 0$, zero-energy state no longer exists such that the energy
of the existing state is always finite ($E=\Phi$ in the limit of
$d=0$). This leads to the splitting of the ZBCP. When $J$ is also
finite, there will be further effect caused by spin polarization
although the splitting peak remains at $E=\Phi$ in the limit of
$d=0$. It is interesting to note that beyond the quasiclassical
approximation, a more accurate calculation for the surface
bound-state energies in $d_{x^{2}-y^{2}}$-wave and other
unconventional cuprate superconductors was reported by Walker {\em
et al.} \cite{Walker97}.

\subsection{Effect of Fermi-Wave-Vector Mismatch}

\begin{figure}[ptb]
\begin{center}
\vspace{-0.5cm}
\includegraphics[width=9.0cm ]{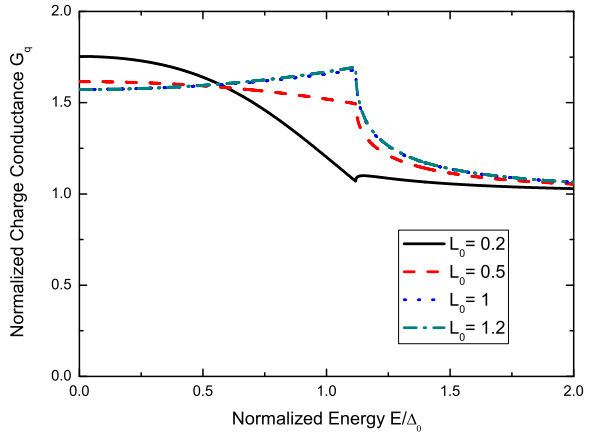}
\end{center}
\vspace{-1.0cm} \caption{Effect of FWM on normalized charge
conductance spectra $G_q$ for various wave-vector mismatch value
$L_0$ with fixed barrier $Z=0$, AF order $\Phi=0.5\Delta_0$, and
spin polarization $X=0.5$.} \label{m3}
\end{figure}

\begin{figure}[ptb]
\begin{center}
\vspace{-0.5cm}
\includegraphics[width=9.0cm ]{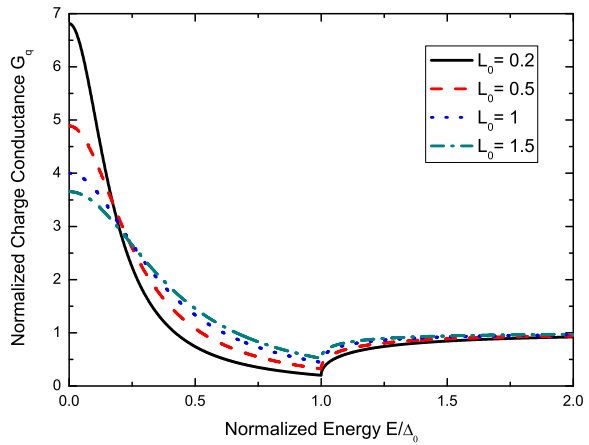}
\end{center}
\vspace{-1.0cm} \caption{Effect of FWM on normalized charge
conductance spectra $G_q$ for various wave-vector mismatch value
$L_0$ with fixed barrier $Z=1$, AF order $\Phi=0$, and spin
polarization $X=0$. This can be considered as the case of
hole-doped cuprate superconductors without AF order and in the
absence of spin polarization.} \label{m4}
\end{figure}

\begin{figure}[ptb]
\begin{center}
\vspace{-0.5cm}
\includegraphics[width=9.0cm ]{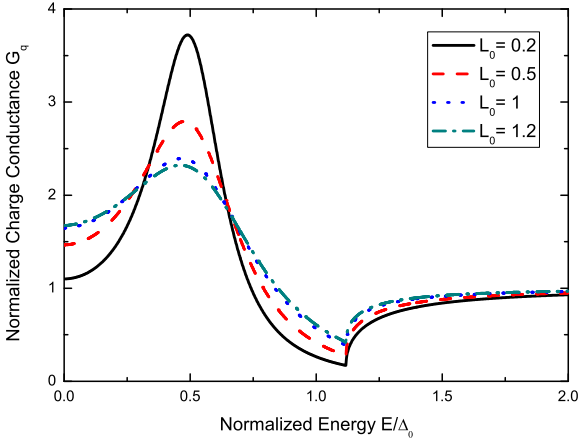}
\end{center}
\vspace{-1.0cm} \caption{Effect of FWM on normalized charge
conductance spectra $G_q$ for various wave-vector mismatch value
$L_0$ with fixed barrier $Z=1$, AF order $\Phi=0.5\Delta_0$, and
spin polarization $X=0.5$. FWM causes the reduction of conductance
at zero bias, while enhances the splitting peak associated with
the AF order.} \label{m5}
\end{figure}

Tunneling conductances are in general strongly modified by the
effect of Fermi-wave-vector mismatch (FWM)
\cite{Zutic00,Linder07}. In our case, due to the presence of the
AF order, the conductance spectra are somewhat different from
those obtained by \ifmmode \check{Z}\else \v{Z}\fi{}uti\ifmmode
\acute{c}\else \'{c}\fi{} and Valls \cite{Zutic00} and Linder {\em
et al.} \cite{Linder07}. Here we introduce a parameter
\begin{eqnarray}
L_0\equiv {q_F \over k_F} \label{eq:L0}
\end{eqnarray}
to account for the effect of FWM. Both $L_0$ greater and smaller
than one cases are considered. As shown in Figs.~\ref{m3}-\ref{m5}
for the normalized charge conductance $G_q$, the effect of FWM is
typically strong when $L_0<1$, while it has relatively minor
effect when $L_0>1$ (That is, $G_q$ changes little from the no FWM
$L_0=1$ case.). As first pointed out by Blonder and Tinkham
\cite{Blonder83}, FWM can be interpreted as a type of barrier
which could enhance the conductance near zero bias.

In our previous paper \cite{Liu07}, it was shown that ZBCP of a
$d_{x^{2}-y^{2}}$-wave superconductor can be split by the AF order
$\Phi$. No spin-active barrier \cite{Kashiwaya99,Linder07},
external magnetic field, and spin polarization effects were
considered in our previous case though. Previously \ifmmode
\check{Z}\else \v{Z}\fi{}uti\ifmmode \acute{c}\else \'{c}\fi{} and
Valls \cite{Zutic00} had given a detailed analysis of the FWM
effect on the conductance in ferromagnet/s-wave and d-wave
superconductor junctions. Here we show how FWM influences the
conductance in the current case and point out the key physics.
Fig.~\ref{m3} plots $G_q$ for various $L_0$ with barrier $Z=0$, AF
order $\Phi=0.5\Delta_0$, and spin polarization $X=0.5$ [see
Eq.~(\ref{eq:X}) for the definition of $X$]. One sees that the
effect of FWM is most noticeable at large FWM ($L_0=0.2$ case) to
which a ZBCP is developed, while the spectra are humdrum when
$L_0\geq 1$. Since no barrier ($Z=0$) is considered, no effect of
AF order and spin polarization is seen in terms of peak splitting.
Note that normalized zero-bias conductance is not equal to 2 due
to the presence of AF order and spin polarization. In order to
compare with the case of hole-doped high-$T_c$ superconductors
(without AF order), Fig.~\ref{m4} plots $G_q$ at different values
of $L_0$ with $\Phi=X=0$ and $Z=1$. One sees that ZBCP is largely
enhanced by the FWM effect (see $L_0=0.2$ case). Thus FWM can
significantly enhance the number of midgap surface states near
zero-bias voltage.

Aiming to electron-doped cuprate superconductors, Fig.~\ref{m5}
shows the effect of FWM on the splitting peak when the AF order is
present ($\Phi=0.5\Delta_0$). Here barrier $Z=1$ and spin
polarization $X=0.5$. In contrast to the case of $\Phi=0$ in
Fig.~\ref{m4}, FWM actually reduces the number of midgap surface
states near zero bias. At the same time, it enhances the strength
of the splitting peak associated with the AF order. Following the
idea of Blonder and Tinkham \cite{Blonder83} such that FWM can be
interpreted as a type of barrier, the enhancement of ZBCP in
Fig.~\ref{m4} and the reduction of zero-bias conductance in
Fig.~\ref{m5} is a natural outcome at large FWM.

In principle, the effect of FWM should be included when a serious
calculation is performed for spin-polarized conductances.

\subsection{Effect of Spin Polarization}

In the literature, there exists different definitions of spin
polarization. One example is the ``tunneling polarization"
proposed by Tedrow and Meservey \cite{Tedrow94}. In point contact
experiment, the more suitable definition is the so-called
``contact polarization" \cite{Soulen98}
\begin{eqnarray}
P_c  = \frac{{N_ \uparrow  (E_F )v_{F \uparrow }  - N_ \downarrow
(E_F )v_{F \downarrow } }}{{N_ \uparrow  (E_F )v_{F \uparrow }  +
N_ \downarrow  (E_F )v_{F \downarrow } }},
\label{eq:pc1}
\end{eqnarray} where $v_{F\sigma}$ and $N_\sigma(E_F)$ are respectively
the Fermi velocity and DOS at Fermi level for spin-$\sigma$
electron. Since $I_\sigma \propto N_\sigma (E_F)v_{F\sigma}$,
Eq.~(\ref{eq:pc1}) is identical to
\begin{eqnarray}
P_c  = \frac{{I_ \uparrow   - I_ \downarrow  }}{{I_ \uparrow   +
I_ \downarrow  }}. \label{eq:pc2}
\end{eqnarray}
However, the most natural definition of spin polarization is
\begin{eqnarray}
X \equiv \frac{{N_ \uparrow  (E_F ) - N_ \downarrow  (E_F )}}{{N_
\uparrow  (E_F ) + N_ \downarrow  (E_F )}}. \label{eq:X}
\end{eqnarray}
In ballistic point contact situation, the electron density of
states in the presence of an exchange field can be written as
$N_\sigma (E_F ) = q_{F\sigma }^2 A/4\pi$, where $A$ is the area
of the interface. Thus $X=J/E_{F}^F$ with $ E_{F}^F\equiv \hbar ^2
q_{F}^2 /2m = (\hbar ^2 q_{F \uparrow }^2 /2m + \hbar ^2 q_{F
\downarrow }^2 /2m)/2$ \cite{Chalsani07}. In Sec.~III.E, we will
show that spin polarization $X$ can be determined by a general
formula in combination with the experimental conductance data.


\begin{figure}[ptb]
\begin{center}
\vspace{-0.5cm}
\includegraphics[width=9.0cm ]{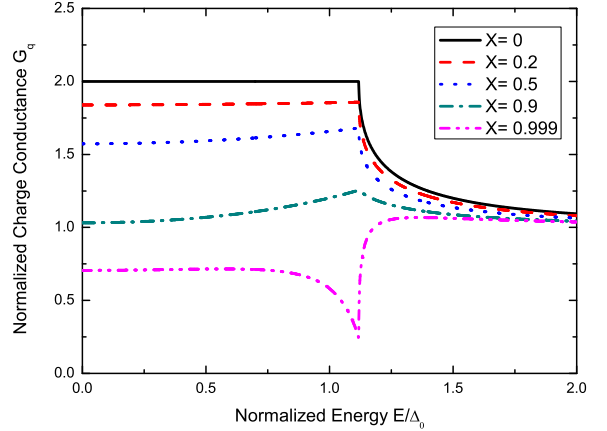}
\end{center}
\vspace{-1.0cm} \caption{Effect of spin polarization on normalized
charge conductance spectra $G_q$ for various spin polarization
value $X$ with fixed barrier $Z=0$, AF order $\Phi=0.5\Delta_0$,
and without FWM ($L_0=1$).} \label{sp2}
\end{figure}

\begin{figure}[ptb]
\begin{center}
\vspace{-0.5cm}
\includegraphics[width=9.0cm ]{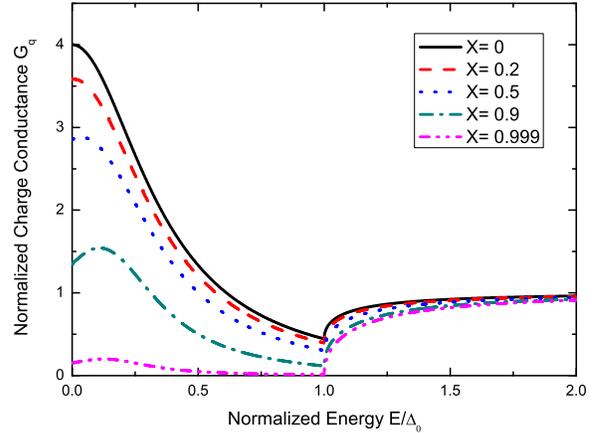}
\end{center}
\vspace{-1.0cm} \caption{Effect of spin polarization on normalized
charge conductance spectra $G_q$ for various spin polarization
value $X$ with fixed barrier $Z=1$, AF order $\Phi=0$, and without
FWM ($L_0=1$). This is considered an example of the hole-doped
cuprate superconductor without AF order and FWM.} \label{sp3}
\end{figure}

\begin{figure}[ptb]
\begin{center}
\vspace{-0.5cm}
\includegraphics[width=9.0cm ]{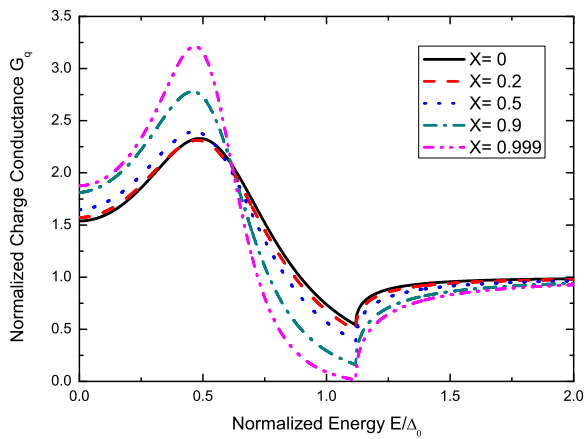}
\end{center}
\vspace{-1.0cm} \caption{Effect of spin polarization on normalized
charge conductance spectra $G_q$ for various spin polarization
value $X$ with fixed barrier $Z=1$, AF order $\Phi=0.5\Delta_0$,
and without FWM ($L_0=1$). Low-energy anomalous conductance
enhancement arises due to AF contributions (see text for
details).} \label{sp4}
\end{figure}

Note that current quasiparticle wave function of BdG equations has
four components which involve two components associated with the
AF order. In the limit of $Z=0$ and without spin polarization
($X=0$), normalized charge conductance has value 2 as expected
(see Fig.~\ref{sp2}). With a finite AF order ($\Phi=0.5\Delta_0$),
the resulting effective gap magnitude is about
$\tilde{\Delta}\simeq 1.12\Delta_0$ (see Fig.~\ref{sp2}). In
general, at $E < \tilde{\Delta}$, effect of spin polarization is
to suppress the conductance. When FWM is absent ($L_0=1$) together
with $Z=0$, normal reflection has no contribution and Andreev
reflection actually dominates the tunneling process for $E <
\tilde{\Delta}$ \cite{Blonder82}. In our current case, Andreev
reflection involves contributions from both $R_A$ and $R_A^{AF}$
channels.

The most interesting results occur when the barrier $Z$ is finite.
When the AF order $\Phi=0$ (as for the case of hole-doped cuprate
superconductors) to which $R_N^{AF}=R_A^{AF}=0$, ZBCP appears
whose (normalized) strength is largely suppressed due to the
strong spin polarization effect (see Fig.~\ref{sp3}). However, as
seen in Fig.~\ref{sp4}, when AF order is finite
($\Phi=0.5\Delta_0$), in contrast, the strengths of both the
zero-bias conductance and the splitting peak turn out to get
enhanced by the strong spin polarization effect. This ``anomalous
conductance enhancement" phenomenon is in drastic contrast as
compared to the ZBCP associated with $\Phi=0$ case
(Fig.~\ref{sp3}). These somewhat surprising results arise due to a
significant increase of $|R_N^{AF}|$ and at the same time, a
significant decrease of $|R_A^{AF}|$ for large $X$ cases -- a
consequence of the interplay between AF order and spin
polarization. Since $|R_N^{AF}|$ contributes positively to the
conductance, while $|R_A^{AF}|$ contributes negatively to the
conductance [see Eq.~(\ref{eq:Cq})], resultantly they cause the
anomalous conductance enhancement at low energies ($E\leq\Phi$).
It should be emphasized that this low-energy conductance
enhancement is not due to the spin-flip effect which is not
considered in this paper. At higher energies, $E>\Phi$, the
conductances behave more normally such that they get suppressed
due to the spin polarization effect. Anomalous conductance
enhancement at low energies can serve as a test to see whether
there is an significant AF order in electron-doped cuprate
superconductors.

Interface barrier and band structure are in general having strong
effect on spin polarization. Kant {\em et al.} have built
an``extended interface" model to illustrate the decay of spin
polarization \cite{Kant02}. Besides, Mazin had a detailed
discussion on the definition of spin polarization and band
structure effects in spin polarization \cite{Mazin99}.


\subsection{Effect of Effective Barrier}

In the study of the tunneling transition in Cu-Nb point contacts,
Blonder and Tinkham \cite{Blonder83} pointed out that barrier is
not the only source for normal reflection and in a more realistic
system, one should consider ``impedance" or FWM as well which
results in normal reflection even with no barrier present. They
proposed an effective barrier $Z_{\rm eff}=[Z^2+(1-r)^2/4r]^{1/2}$
where $r$ is the Fermi velocity ratio. They showed that effective
barrier has an obvious effect on the conductance when
$E<\Delta_0$, as shown in Fig.~2 of Ref.~\cite{Blonder83}. Here we
generalize their idea to consider a spin, FWM, and angle dependent
effective barrier $Z_{\rm eff}$ \cite{Blonder83,Zutic00}:
\begin{eqnarray}
Z_{\rm eff}  \equiv  [Z^2  + (1 - L_\sigma  )^2 /4L_\sigma
]^{1/2}/ \cos \theta _{S\sigma }, \label{eq:zeff}
\end{eqnarray}
where $L_\sigma=q_{F\sigma } /k_F$ corresponds to the
spin-dependent FWM. It is noted that we are not considering the
spin-active barrier which has spin filtering effects and can lead
to the ZBCP splitting \cite{Kashiwaya99,Linder07}. Instead we
propose a possible alternative mechanism to account for the decay
of spin polarization. Based on the generalized effective barrier,
spin-up and -down particles experience different strength of
effective barrier that causes spin-up and -down currents to
decrease at different speed as compared to the current in the
absence of barrier. Consequently, $Z_{\rm eff}$ can modify the
values of $I_ \uparrow - I_ \downarrow$ (and thus $P_c$)
dramatically. With this strong effect at work, the decay of spin
polarization should not be dominant by the spin-flitting process
in the point contact spin polarization case.

\begin{figure}[ptb]
\begin{center}
\vspace{-0.5cm}
\includegraphics[width=9.0cm ]{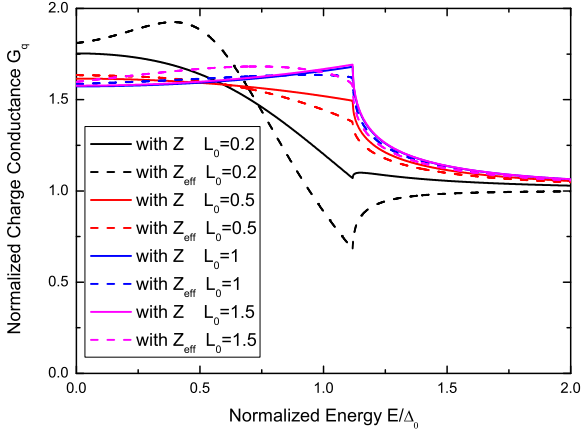}
\end{center}
\vspace{-1.0cm} \caption{Effect of effective barrier $Z_{\rm eff}$
on normalized charge conductance spectra $G_q$ for various values
of FWM $L_0$. AF order $\Phi=0.5\Delta_0$ and spin polarization
$X=0.5$. The bare barrier $Z$ is set to zero, while $Z_{\rm eff}$
is given by Eq.~(\ref{eq:zeff}).} \label{barrier1}
\end{figure}


As seen in Eq.~(\ref{eq:zeff}), $Z_{\rm eff}$ can differ
significantly from $Z$, especially when $Z$ is small. Essentially
their difference can be measured by spin-polarized tunneling
experiments. In Fig.~\ref{barrier1}, we compare the effects of $Z$
and $Z_{\rm eff}$ on the conductance with bare barrier $Z$ set to
zero and vary the FWM $L_0$ value. For $Z=0$ and $\theta _{S\sigma
}=0$, $Z_{\rm eff}=[(1 - L_\sigma )^2 /4L_\sigma ]^{1/2}$ [see
(\ref{eq:zeff})]. In our case, we have also included AF order and
spin polarization. The difference is most noticeable when FWM is
large ($L_0=0.2$ case). Since $Z=0$, AF order and spin
polarization have little effect at small FWM. However, when FWM is
large, AF order and spin polarization can have a strong effect
such that a splitting peak can develop at $E\approx
\Phi=0.5\Delta_0$ with the effective barrier $Z_{\rm eff}$ (see
Fig.~\ref{barrier1}). This supports Blonder and Tinkham's idea of
``impedance" mismatch which enhances the normal reflection.

\subsection{A General Formula for Determining the Spin Polarization}

Based on the phenomenon of Andreev reflection, Soulen {\em et al.}
\cite{Soulen98} proposed a formula for determining the point
contact spin polarization $P_c$ [see Eqs.~(\ref{eq:pc1}) and
(\ref{eq:pc2})] when the normalized zero-bias conductance data is
compared. Their original form was
\begin{eqnarray}
G(0)/G_N  = 2(1 - P_c), \label{ratio}
\end{eqnarray}
which is valid only when FWM is absent \cite{Strijkers01}. Since
Andreev reflection could be strongly modified due to the FWM
effect, it's useful to replace Eq.~(\ref{ratio}) by
\begin{eqnarray}
G(0)/G_N  = \left[1 + \left| {R_A } \right|^2  - \left| {R_A^{AF}
} \right|^2 \right](1 - P_c), \label{f1}
\end{eqnarray}
where $R_A$ and $R^{AF}_A$ are the AR and AF-AR coefficients
respectively. Eq.~(\ref{f1}) can be reduced to Eq.~(\ref{ratio})
when the exchange energy $J$ is set to zero in $R_A$ and the AF
order $\Phi$ is set to zero in $R^{AF}_A$. Note also that the
parameter $X$ should be set to zero when the ``contact
polarization" $P_c$ is determined under the idea of Soulen {\em et
al}.

Here we propose a more general formula for determining the spin
polarization:
\begin{eqnarray}
G(0)/G_{N}  = A_ \uparrow   + A_ \downarrow, \label{general}
\end{eqnarray} where
\begin{eqnarray}
A_ \uparrow = \int_\alpha ^\beta d\theta _{N\sigma} \cos \theta
_{N\sigma}(1 + a_\uparrow| {R_{A \downarrow } }|^2 -a_\uparrow|
{R_{A\downarrow }^{AF}}|^2)P_ \uparrow
\end{eqnarray}
and
\begin{eqnarray}
A_ \downarrow = \int_\alpha ^\beta  {d\theta _{N\sigma} \cos
\theta _{N\sigma} (1+|{R_{A \uparrow }}|^2  -| {R_{A\uparrow
}^{AF}}|^2 )} P_ \downarrow.
\end{eqnarray}
Here $R_{A\sigma }  = R_{A\sigma } (L_0 ,X,\Phi ,\theta _{N\sigma}
)$ and $R_{A\sigma }^{AF}  = R_{A\sigma }^{AF} (L_0 ,X,\Phi
,\theta _{N\sigma} ) $ with $E=0$. Eq.~(\ref{general}) is a
natural result of our earlier formalism. It is regarded as the
generalization of Eq.~(\ref{ratio}) of Soulen {\em et al.}, which
includes the effects of FWM, spin polarization, AF order, as well
as the incident angle.

\section{Conclusions}
Tunneling experiment provides a useful tool for probing the
properties of a superconductor such as the magnitude and symmetry
of the superconducting order parameter, quasiparticle density of
states, and any existing competing orders. In fact, tunneling
experiment is also a powerful probe for investigating the
spin-charge separation in connection with the spin-injection
techniques. This involves both charge imbalance and spin imbalance
studies.

In this paper, we have presented a detailed study of the tunneling
conductance spectra of a ferromagnetic metal/electron-doped
superconductor junctions, taking into account an AF order existing
in the the electron-doped superconductor. Interesting result, such
as low-energy anomalous conductance enhancement, occurs as a
result of the interplay between AF order and spin polarization
(see Fig.~\ref{sp4}). These results in turn provide a further
opportunity to test whether there is an significant AF order in
electron-doped cuprate superconductors.

\acknowledgements

This work is supported by National Science Council of Taiwan
(Grant No. 96-2112-M-003-008) and National Natural Science
Foundation of China (Grant No. 10347149). We also acknowledge the
support from the National Center for Theoretical Sciences, Taiwan.

\appendix
\section{Reflection Coefficients}
\label{appendix:A}
Under the WKBJ approximation
~\cite{Bardeen69,Bar-Sagi72,Hu75,Bruder90,Blonder82,Hu94,Tanaka95,Kashiwaya96},
the wave functions in the generalized BdG equations (\ref{eq:BdG})
can be approximated by
\begin{eqnarray}
\left( {\begin{array}{*{20}c}
   {u_{1\sigma} }  \\
   {v_{1\bar\sigma} }  \\
   {u_{2\sigma} }  \\
   {v_{2\bar\sigma} }  \\
\end{array}} \right) = \left( {\begin{array}{*{20}c}
   {e^{i{\bf{k}}_F  \cdot {\bf{r}}} \tilde u_{1\sigma} }  \\
   {e^{i{\bf{k}}_F  \cdot {\bf{r}}} \tilde v_{1\bar\sigma} }  \\
   {e^{i{\bf{k}}_{F{\bf Q}}  \cdot {\bf{r}}} \tilde u_{2\sigma} }  \\
   {e^{i{\bf{k}}_{F{\bf Q}}  \cdot {\bf{r}}} \tilde v_{2\bar\sigma} }  \\
\end{array}} \right).
\end{eqnarray}
Thus one obtains a set of Andreev equations in the $x$ direction,

\begin{eqnarray}
E  \tilde u_{1\sigma} (x) &=& H_\sigma  \tilde u_{1\sigma} (x) +
 \Delta ({\bf{\hat k}}_F )\tilde v_{1\bar \sigma} (x) + \Phi \tilde u_{2\sigma} (x)
 \nonumber\\
E  \tilde v_{1\bar \sigma} (x) &=& \Delta ^ *  ({\bf{\hat k}}_F
)\tilde u_{1\sigma} (x) -
 H_\sigma  \tilde v_{1\bar \sigma} (x) + \Phi \tilde v_{2\bar \sigma} (x) \nonumber\\
 E  \tilde u_{2\sigma} (x) &=& \Phi \tilde u_{1\sigma} (x) - H_\sigma  \tilde u_{2\sigma} (x) +
 \Delta ({\bf{\hat k}}_{F{\bf Q}} )\tilde v_{2\bar \sigma} (x)
 \nonumber\\
 E  \tilde v_{2\bar \sigma} (x) &=& \Phi \tilde v_{1\bar \sigma} (x) + \Delta ^ *
 ({\bf{\hat k}}_{F{\bf Q}} )\tilde u_{2\sigma} (x) + H_\sigma  \tilde v_{2\bar \sigma} (x) ,
 \nonumber\\
 \label{eq:BdG2}
\end{eqnarray} where $H_\sigma   =  - \frac{{i\hbar
^2 k_F }}{m}\frac{d}{{dx}} - \sigma J$ and $x$ is the coordinate
normal to the interface. The $d_{x^2-y^2}$-wave SC gap $\Delta
({\bf{\hat k}}_F )=-\Delta({\bf{\hat k}}_{F{\bf Q}})\equiv \Delta
_{0}\sin 2\theta$ with $\theta$ the azimuthal angle relative to
the $x$-axis. In obtaining Eq.~(\ref{eq:BdG2}), the Fourier
transform of the Cooper pair order parameter $\Delta({\bf s},{\bf
r})$ from relative coordinate ${\bf s}$  to ${\bf k}$ space is
assumed to take the form, $\Delta ({\bf{k}},{\bf{r}}) = \Delta
({\bf\hat k}_F)\Theta (x)$, with $\Theta \left( x\right) $ the
Heaviside step function \cite{Hu94,Kashiwaya96}.

Solving Eq.~(\ref{eq:BdG2}), one obtains four eigenvectors which
build up the spin-$\sigma$ wave function in the superconductor
region ($x>0$) \cite{Zutic00},
\begin{eqnarray}
 \psi _{S\sigma} (x) = \left[ {c_{1\sigma} \left( {\begin{array}{*{20}c}
   \Delta   \\
   {E_{  - } }  \\
   0  \\
   \Phi   \\
\end{array}} \right) + c_{2\sigma} \left( {\begin{array}{*{20}c}
   {E_{  + } }  \\
   \Delta   \\
   \Phi   \\
   0  \\
\end{array}} \right)} \right] e^{ik^ +  x}\nonumber\\
  + \left[ {c_{3\sigma} \left( {\begin{array}{*{20}c}
   {E_{  - } }  \\
   { - \Delta }  \\
   \Phi   \\
   0  \\
\end{array}} \right) + c_{4\sigma} \left( {\begin{array}{*{20}c}
   { - \Delta }  \\
   {E_{  + } }  \\
   0  \\
   \Phi   \\
\end{array}} \right)} \right] e^{ - ik^ - x}.
\label{eq:phi_s1}
\end{eqnarray} Here $E_{ \pm } \equiv E \pm
\varepsilon_\sigma $ with $\varepsilon_\sigma = \sqrt {E^2 -
\Delta ^2 - \Phi ^2 }$, $\Delta\equiv\Delta({\bf{\hat k}}_F )$,
$k^ + =k^ - = k_{F} \cos \theta _{S\sigma}$, and $c_{i\sigma}$ are
coefficients of the corresponding waves. As pointed out by Blonder
{\em et al.} \cite{Blonder82}, there is no need to normalize the
coefficients as it just complicates the calculation. If we set
$\Phi=J=0$ and normalize the coefficients, it will reduce to the
case for a typical N/I/S junction
\cite{Blonder82,Tanaka95,Kashiwaya96}.

Since we consider that there is an AF order in the EDSC side, an
incident electron from the FM side will have four possible
reflections \cite{Liu07}. The spin-$\sigma$ wave function in the FM
side $(x<0)$ with incident angle $\theta _{N\sigma}$ can thus be
written as \cite{Blonder82,Kashiwaya96,Kashiwaya99}
\begin{eqnarray}
\Psi _{N\sigma } (x) = \left( {\begin{array}{*{20}c}
   {e^{iq_{F\sigma } \cos \theta _{N\sigma} x}  + R_{N\sigma } e^{ - iq_{F\sigma }
   \cos \theta _{N\sigma} x} }  \\
   {R_{A\bar \sigma } e^{iq_{F\bar \sigma } \cos \theta _{A\bar \sigma } x} }  \\
   {R^{AF}_{N\sigma } e^{iq_{F\sigma } \cos \theta _{N\sigma} x} }  \\
   {R^{AF}_{A\bar \sigma } e^{ - iq_{F\bar \sigma } \cos \theta _{A\bar \sigma } x} } \\
\end{array}} \right),
\label{state at metal}
\end{eqnarray}
where $R_{N\sigma }$, $R_{A\bar \sigma }$, $R^{AF}_{N\sigma }$, and
$R^{AF}_{A\bar \sigma }$ are amplitudes of NR, AR, AF-NR, and AF-AR
respectively. Applying the following boundary conditions:
\begin{eqnarray}
\psi_{N\sigma }\left( x\right) |_{x=0^{-}}& =&\psi _{S\sigma}\left(
x\right) |_{x=0^{+}}
\label{eq:boundary conditions} \\
\frac{2mH}{\hbar ^{2}}\psi _{S\sigma}\left( x\right) |_{x=0^{+}}&
=&\frac{d\psi
_{S\sigma}\left( x\right) }{dx}|_{x=0^{+}}-\frac{d\psi _{N\sigma }\left( x\right) }{dx}%
|_{x=0^{-}}, \nonumber
\end{eqnarray}
the four reflection amplitudes (coefficients) are solved to be
\begin{eqnarray}
 R_{N\sigma }&=& \frac{{E_{-}(1 - L_\sigma  \lambda _{1\sigma }
 + 2iZ_\theta)B}}{{(1 + L_\sigma  \lambda _{1\sigma }  +
 2iZ_\theta)D}}\nonumber\\
  &-& \frac{{\Delta (1 + L_{\bar \sigma } \lambda _{2\bar \sigma }
   + 2iZ_\theta)A}}{{(1 + L_\sigma  \lambda _{1\sigma }  + 2iZ_\theta)D}}\nonumber\\
   &-&  \frac{{1 - L_\sigma  \lambda _{1\sigma }  - 2iZ_\theta}}
   {{1 + L_\sigma  \lambda _{1\sigma }  + 2iZ_\theta}}\nonumber \\
 R_{A\bar \sigma } &=& \frac{{\Delta (1 + L_\sigma  \lambda _{1\sigma }
 - 2iZ_\theta)B}}{{(1 + L_{\bar \sigma } \lambda _{2\bar \sigma }
 - 2iZ_\theta)D}}\nonumber\\
 &+&  \frac{{E_{-}(1 - L_{\bar \sigma } \lambda _{2\bar \sigma }
  - 2iZ_\theta)A}}{{(1 + L_{\bar \sigma } \lambda _{2\bar \sigma }  - 2iZ_\theta)D}} \nonumber\\
 R^{AF}_{N\sigma }&=& \frac{{\Phi B}}{D} \nonumber\\
 R^{AF}_{A\bar \sigma }&=& \frac{{\Phi A}}{D},
\label{eq:R}
\end{eqnarray} where
\begin{eqnarray}
 A &=& 2\Delta L_\sigma  \lambda _{1\sigma } [1 - L_\sigma  L_{\bar \sigma }
 \lambda _{1\sigma } \lambda _{2\bar \sigma }  + 4Z_\theta^2\nonumber\\
 &+& 2iZ_\theta(L_\sigma  \lambda _{1\sigma }
 + L_{\bar \sigma } \lambda _{2\bar \sigma } )] \nonumber\\
 B &=& 2L_\sigma  \lambda _{1\sigma } [2L_{\bar \sigma } \lambda _{2\bar \sigma } E
 + \varepsilon (1 + L_{\bar \sigma }^2 \lambda _{2\bar \sigma }^2 )] \nonumber\\
 D &=& \Delta ^2 [(1 - L_\sigma  L_{\bar \sigma } \lambda _{1\sigma }
 \lambda _{2\bar \sigma }  + 4Z_\theta^2 )^2 \nonumber\\
 &+& 4Z_\theta^2 (L_\sigma  \lambda _{1\sigma }
 + L_{\bar \sigma } \lambda _{2\bar \sigma } )^2 ] \nonumber\\
 &+& [2L_\sigma  \lambda _{1\sigma } E + 4\varepsilon Z_\theta^2
 + \varepsilon (1 + L_\sigma ^2 \lambda _{1\sigma }^2 )]\nonumber\\
 &\times&[2L_{\bar \sigma } \lambda _{2\bar \sigma } E
 + 4\varepsilon Z_\theta^2  + \varepsilon (1 + L_{\bar \sigma }^2
 \lambda _{2\bar \sigma }^2 )].
 \end{eqnarray}
Moreover $ Z_\theta = Z /\cos \theta _{S\sigma } $ with the barrier
$Z=mH/\hbar ^2 k_{F}$, $\lambda _{1\sigma }  = \cos \theta
_{N\sigma} /\cos \theta _{S\sigma } $, $ \lambda _{2\bar \sigma } =
\cos \theta _{A\bar \sigma } /\cos \theta _{S\sigma }$, and ${\rm{
}}L_\sigma =\sqrt {q_F /k_F - \sigma (q_F /k_F )(J/E_{FN} )}$. It is
interesting to note in (\ref{eq:R}) that $R^{AF}_{N\sigma }$ and
$R^{AF}_{A\bar \sigma }$ are proportional to the AF order $\Phi$, as
is expected.

\section{Midgap Surface States}

\label{appendix:B} Following Ref.~\cite{Hu94}, we first assume
that
\begin{eqnarray}
\left(
\begin{array}{c}
\tilde{u}_{l\sigma} \\
\tilde{v}_{l\sigma}%
\end{array}%
\right) =e^{-\gamma_{\sigma} x}\left(
\begin{array}{c}
\hat{u}_{l\sigma} \\
\hat{v}_{l\sigma}%
\end{array}%
\right), \label{eq:uv_gamma}
\end{eqnarray} where $\gamma_\sigma $
is the attenuation constant for $| {E({\bf{q}}_{F\sigma } )}| <
\sqrt {| {\Delta ({\bf{\hat k}}_F )} |^2  + \Phi ^2 } $. With
(\ref{eq:uv_gamma}), Eq.~(\ref{eq:BdG2}) becomes
\begin{eqnarray} E\left( {\begin{array}{*{20}c}
   {\hat u_{1\sigma} }  \\
   {\hat v_{1\bar\sigma} }  \\
   {\hat u_{2\sigma} }  \\
   {\hat v_{2\bar\sigma} }  \\
\end{array}} \right) = \left( {\begin{array}{*{20}c}
   {h} & {\Delta} & \Phi  & 0  \\
   {\Delta} & { - h} & 0 & \Phi   \\
   \Phi  & 0 & { - h} & { - \Delta}  \\
   0 & \Phi  & { - \Delta} & {h}  \\
\end{array}} \right)\left( {\begin{array}{*{20}c}
   {\hat u_{1\sigma} }  \\
   {\hat v_{1\bar\sigma} }  \\
   {\hat u_{2\sigma} }  \\
   {\hat v_{2\bar\sigma} }  \\
\end{array}} \right)
\label{Andereev equations}
\end{eqnarray}
for the superconducting overlayer ($x>0$). Here $ h  =
{\varepsilon^\prime_\sigma} - \sigma J $ with $
{\varepsilon^\prime_\sigma} = i\hbar ^2 m^{ - 1} \gamma_\sigma q_F
\cos \theta _{N\sigma} $. The wave-vector components parallel to the
interface are conserved for all possible processes.

Solving Eq.~(\ref{Andereev equations}), one obtains double
degenerate eigenvalues $E =  \pm \sqrt {\Delta ^2  + \Phi ^2  +
{\varepsilon^\prime_\sigma}^2 }  - \sigma J $, where $+$ ($-$)
corresponds to the electron- (hole-) like QP excitation. Similar to
the wave function (\ref{eq:phi_s1}), superposition of the four
eigenstates makes up the formal wave function for the superconductor
overlayer ($x>0$)
\begin{eqnarray}
 \psi _{S\sigma} (x) = \left[ {c_{1\sigma} \left( {\begin{array}{*{20}c}
   \Delta   \\
   {E_{  - } }  \\
   0  \\
   \Phi   \\
\end{array}} \right) + c_{2\sigma} \left( {\begin{array}{*{20}c}
   {E_{  + } }  \\
   \Delta   \\
   \Phi   \\
   0  \\
\end{array}} \right)} \right]e^{ - \gamma_\sigma x} e^{ik^ +  x}\nonumber\\
  + \left[ {c_{3\sigma} \left( {\begin{array}{*{20}c}
   {E_{  - } }  \\
   { - \Delta }  \\
   \Phi   \\
   0  \\
\end{array}} \right) + c_{4\sigma} \left( {\begin{array}{*{20}c}
   { - \Delta }  \\
   {E_{  + } }  \\
   0  \\
   \Phi   \\
\end{array}} \right)} \right]e^{ - \gamma_\sigma x} e^{ - ik^ -  x}. \nonumber \\
\label{state at SU overlayer}
\end{eqnarray}
Here $E_{ \pm } \equiv E \pm \varepsilon^\prime_\sigma$ with $
\varepsilon^\prime_\sigma   = \sqrt {(E + \sigma J)^2  - \Delta ^2
- \Phi ^2 } $, $c_{i}$ are coefficients of the corresponding
waves, and $k^ + =k^ -  = k_{F} \cos \theta _{S\sigma}$. At the
interface, the wave functions of FM and superconductor meet ideal
continuity $\psi _{N\sigma}(x=0)=\psi _{S\sigma}(x=0)$. After some
algebra, the formal wave function for the FM overlayer is obtained
to be ($x<0$):

\begin{eqnarray}
 \psi _{N\sigma } (x) =\left[ {c_{1\sigma} \left( {\begin{array}{*{20}c}
   {e^{ik_{1\sigma} x} \Delta}  \\
   {e^{ - ik_{1\sigma} x} E_{  - } }  \\
   0  \\
   {e^{ik_{1\sigma} x} \Phi } \nonumber \\
\end{array}} \right) + c_{2\sigma} \left( {\begin{array}{*{20}c}
   {e^{ik_{1\sigma} x} E_{  + } }  \\
   {e^{ - ik_{1\sigma} x} \Delta}  \\
   {e^{ - ik_{1\sigma} x} \Phi }  \nonumber \\
   0  \\
\end{array}} \right)} \right]e^{ik^ +   x}  \\
  + \left[ {c_{3\sigma} \left( {\begin{array}{*{20}c}
   {e^{ - ik_{1\sigma} x} E_{  - } }  \\
   { - e^{ik_{1\sigma} x} \Delta}  \\
   {e^{ik_{1\sigma} x} \Phi } \nonumber \\
   0  \\
\end{array}} \right) + c_{4\sigma} \left( {\begin{array}{*{20}c}
   { - e^{ - ik_{1\sigma} x} \Delta}  \\
   {e^{ik_{1\sigma} x} E_{  + } }  \\
   0  \\
   {e^{ - ik_{1\sigma} x} \Phi }  \\
\end{array}} \right)} \right]e^{ - ik^ -   x},  \\
\label{state at FM overlayer}
\end{eqnarray}
where it is assumed that incident spin-$\sigma$ electron has the
wave vector $k_{1\sigma}$ along the $x$ direction. Considering the
free boundary at $x=-d$, $\psi _{N\sigma } (x=-d) =0$, one thus
obtains the condition for the surface bound states:
\begin{eqnarray}
e^{ - 2ik_{1\sigma} d} E_{  + }  + e^{2ik_{1\sigma} d} E_{  - }  =
2\Phi. \label{surface bound states}
\end{eqnarray}


\end{document}